\documentclass[conference]{IEEEtran}
\IEEEoverridecommandlockouts
\usepackage{cite}
\usepackage{amsmath,amssymb,amsfonts}
\usepackage{algorithmic}
\usepackage{graphicx}
\usepackage{textcomp}
\usepackage{xcolor}
\usepackage{multirow}
\usepackage{booktabs}
\usepackage{bm}

\def\BibTeX{{\rm B\kern-.05em{\sc i\kern-.025em b}\kern-.08em
    T\kern-.1667em\lower.7ex\hbox{E}\kern-.125emX}}
\begin{document}

\title{The DKU-Duke-Lenovo System Description for \\ the Third DIHARD Speech Diarization Challenge
}

\author{\IEEEauthorblockN{1\textsuperscript{st} Weiqing Wang}
\IEEEauthorblockA{\textit{Department of ECE} \\
\textit{Duke University}\\
Durham, USA \\
weiqing.wang@duke.edu}
\and
\IEEEauthorblockN{2\textsuperscript{nd} Qingjian Lin}
\IEEEauthorblockA{\textit{Lenovo Voice Department} \\
\textit{Lenovo AI Lab}\\
Beijing, China \\
linqj3@lenovo.com}
\and
\IEEEauthorblockN{3\textsuperscript{rd} Danwei Cai}
\IEEEauthorblockA{\textit{Department of ECE} \\
\textit{Duke University}\\
Durham, USA \\
danwei.cai@duke.edu}
\and
\IEEEauthorblockN{4\textsuperscript{th} Lin Yang}
\IEEEauthorblockA{\textit{Lenovo Voice Department} \\
\textit{Lenovo AI Lab}\\
Beijing, China \\
yanglin13@lenovo.com}
\and
\IEEEauthorblockN{5\textsuperscript{th} Ming Li}
\IEEEauthorblockA{\textit{Data Science Research Center} \\
\textit{Duke Kunshan University}\\
 Kunshan, China \\
ming.li369@duke.edu}
}
\maketitle

\begin{abstract}
In this paper, we present the submitted system for the third DIHARD Speech Diarization Challenge from the DKU-Duke-Lenovo team. Our system consists of several modules: voice activity detection (VAD), segmentation, speaker embedding extraction, attentive similarity scoring, agglomerative hierarchical clustering. In addition, the target speaker VAD (TSVAD) is used for the phone call data to further improve the performance. Our final submitted system achieves a DER of 15.43\% for the core evaluation set and 13.39\% for the full evaluation set on task 1, and we also get a DER of 21.63\% for core evaluation set and 18.90\% for full evaluation set on task 2. 
\end{abstract}


\section{Introduction}
Speaker diarization is the task of breaking up the audio into homogeneous pieces that belong to the same speaker, and it aims to determine ``who spoke when'' in a continuous audio recording. A traditional speaker diarization system always contains several modules: voice activity detection (VAD), segmentation, speaker embedding extraction, and clustering. 

While this kind of traditional speaker diarization system has been successful in many domains, including meeting, interview, conversation, it still difficult to mitigate the success to more challenging corpora, such as web videos, speech in the wild, child language recordings, etc. \cite{ryanta2018enhancement}. One of the problems is recognizing the speaker in an overlapping region, and an extremely noisy background is also harmful to a diarization system. To raise more researchers’ attention on such challenging data, the third DIHARD speech diarization challenge was held, where the data is drawn from a diverse sampling of sources \cite{ryant2020third}.

It is difficult for traditional speaker diarization system to find the speaker in an overlapping region, but some pre- and post-processing can be employed to solve this problem. In the VoxCeleb Speaker Recognition Challenge 2020 (VoxSRC-20) \cite{nagrani2020voxsrc}, Xiong et al. \cite{xiao2020microsoft} employed conformer-based continuous speech separation (CSS) as the pre-processing to separate the overlapping speech. Besides, Ivan et al. used target-speaker voice activity detection (TSVAD) as post-processing to predicts the activity of each speaker on each time frame on the CHIME-6 challenge. Recently, an end-to-end system was proposed in \cite{horiguchi2020end}, and it can also recognize the speaker in overlap, which shows a better performance than the traditional method does in the CallHome corpus. 

Our submitted system contains several modules. First, we partition the data into conversation telephone speech (CTS) and non-conversation telephone speech (NCTS) since the CTS data is upsampled to 16k from 8k audio signal. Second, for CTS and NCTS data, we train an 8k and a 16k speaker embedding extractor to extract embeddings for audio segments. Third, we perform different clustering methods on CTS and NCTS data, including agglomerative hierarchical clustering (AHC) and spectral clustering (SC). Finally, we employ TSVAD on the CTS data, which significantly improve the performance. For task 2, an additional ResNet-based VAD model is employed to remove the non-speech region from the data.

The rest of this paper is organized as follows: Section 2 describes the details of the dataset we used in this challenge. Section 3 introduces our submitted systems and algorithms for different tasks. Section 4 presents the experimental results and analysis. Finally, section 5 concludes this paper. 

\section{Data Partition and Data Resources}

From the evaluation plan, we notice that the conversation telephone speech (CTS) data are upsampled from 8kHz audio signal while others are 16kHz audio signal. In addition, the CTS data only contains two speakers. Considering that the CTS data is so different from the remaining non-conversation telephone speech (NCTS) 16kHz audio signal, we build two different systems for CTS data and NCTS data. For NCTS data, we employ the system described in \cite{Lin2020}. For CTS data, we first use AHC to determine the homogeneous speaker region. Then, we extract speaker embedding for each speaker and perform TSVAD to get the diarization results. 

To partition the evaluation set into CTS and NCTS data, we extract the  STFT spectrogram on the first 100 seconds of each recording and compare the maximum value in the frequency bin above 4kHz. If this maximum value is greater than the threshold, the recording is classified as NCTS; otherwise, it is CTS data. The threshold is 0.07, which is obtained from the development set. Finally, we downsample the CTS data to 8kHz.

For NCTS data, we use Voxceleb 1 \& 2 \cite{voxceleb} as the training dataset for speaker embedding extraction. AMI meeting corpus \cite{mccowan2005ami}, ICSI meeting corpus \cite{janin2003icsi} and voxconverse dev set \cite{nagrani2020voxsrc} are used for similarity measurement. MUSAN dataset \cite{musan} is employed for data augmentation. 

For CTS data, we first downsample the Voxceleb 1 \& 2 data to 8kHz and then train another speaker embedding model that is suitable for 8k data. Finally, the TSVAD model is trained on a collection of SRE-databases, including SRE 2004, 2005, 2006, 2008, and Switchboard. 

\begin{figure}
    \centering
    \includegraphics[scale=1]{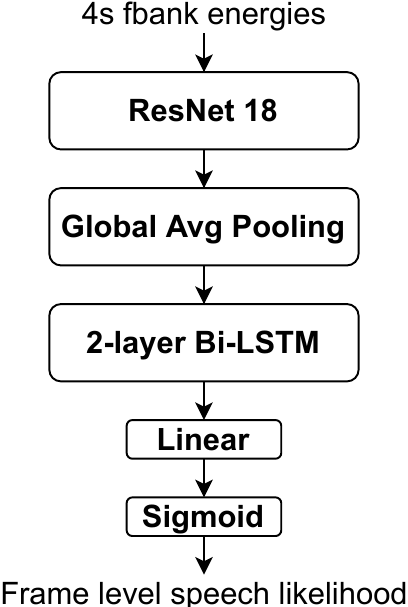}
    \caption{The architecture of the VAD model}
    \label{fig:VAD}
\end{figure}

\section{Detailed Description of Algorithm}

\subsection{VAD}
The architecture of the VAD model is shown in Figure \ref{fig:VAD}. The VAD model consists of a ResNet18 \cite{he2016deep}, a global average pooling (GAP) layer, a 2-layer Bi-LSTM with 64 units per direction as well as a dropout rate of 0.5, and two fully-connected layers followed by a sigmoid function. The widths (number of channels) of the residual blocks are \{16, 32, 64, 128\}, and the corresponding strides are \{(1, 1), (1, 2), (1, 2), (1, 2)\}. The dimensions of two fully-connected layers are 64 and 1, respectively. Given a Mel-filterbank, the ResNet18 can extract the feature maps for speech and non-speech regions. Then, the GAP layer is employed on each channel and get a C-dimensional vector for each frame, where C is the number of channels. Finally, a 2-layer Bi-LSTM captures the sequential information, and a fully-connected layer predicts the frame-level speech likelihood.

The VAD model is trained on the DIHARD III development set, where 90\% of data is for training and the remaining data is for validation. Data augmentation with MUSAN and RIRS corpora is employed to improve the performance, where ambient noise and music are used for the background additive noise and RIRS for reverberation. We do not split the data into CTS and NCTS in VAD model training. 

The acoustic features are 32-dimensional log Mel-filterbank energies with a frame length of 25ms and a hop size of 10ms. The data is broken up into 4s segments with a shift of 2s. During the training phase, we employ the stochastic gradient descent (SGD) optimizer and the binary cross-entropy (BCE) loss with an initial learning rate of 0.1. The learning rate decrease by a factor of 0.1 every 20 epoch. For evaluation, we also break up all data into 4s segments with 2s overlap. The output of the overlapping region between two consecutive segments is the mean of the prediction of these segments. The decision threshold is set to 0.5. 

\subsection{Speaker Embedding Extraction}

We adopt the same structure in \cite{cai2018exploring} as the speaker embedding model, including three components: a front-end pattern extractor, an encoder layer, and a back-end classifier. We employ the ResNet34 as the front-end pattern extractor, where the widths (number of channels) of the residual blocks are \{32, 64, 128, 256\}. Then, a global statistic pooling (GSP) layer projects the variable length input to the fixed-length vector. This vector contains the mean and standard deviation of the output feature maps. Finally, a fully connected layer extracts the 128-dimensional speaker embedding. We use the ArcFace \cite{deng2019arcface} (s=32,m=0.2) as the classifier. The detailed configuration of the neural network is the same as \cite{qin2020ffsvc}. 

We also perform data augmentation with MUSAN and RIRS datasets. For the MUSAN corpus, we use ambient noise, music, television, and babble noise for the background additive noise. For the RIRS corpus, we only use audio from small and medium rooms and perform convolution with training data. 

The acoustic features are 80-dimensional log Mel-filterbank energies with a frame length of 25ms and a hop size of 10ms. The extracted features are mean-normalized before feeding into the deep speaker network. We train two speaker embedding model. One is trained with 16kHz data, which is used for NCTS data. Another is trained with 8kHz downsampled data, which is used in AHC and TSVAD model for CTS data. 


\subsection{Segmentation}
For NCTS data, we employ uniform segmentation with a window length of 1.5s and a shift of 0.75s on the speech region for training, and a window length of 1.5s and a shift of 0.25s on the speech region for inference.

For CTS data, we first employ uniform segmentation with a window length of 0.5s and a shift of 0.25s on the speech region. Then, we extract speaker embedding for each segment and merge the consecutive segments if the cosine similarity of these two segments is greater than a predefined threshold. After two segments are merged to a new segment, the embedding of this new segment becomes the mean of the previous two segments. We merge these segments recursively until all cosine similarity between two consecutive segments is lower than the threshold. The threshold is set to 0.6, which is tuned from the development set.

\begin{figure}
    \centering
    \includegraphics[scale=1]{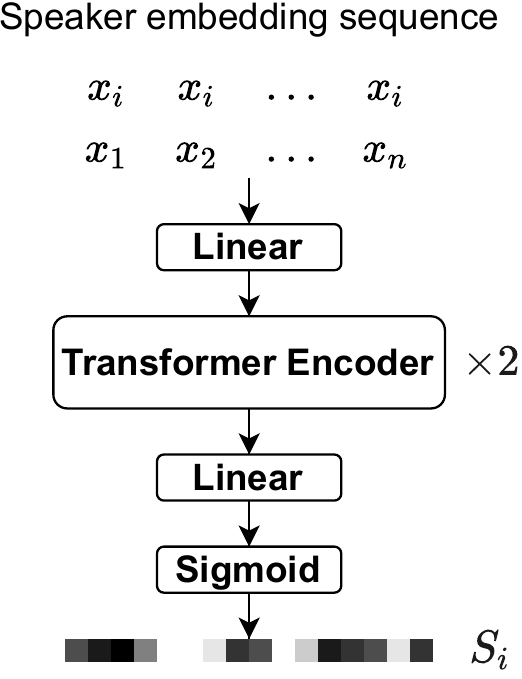}
    \caption{The architecture of the Att-v2s model}
    \label{fig:Att-v2s}
\end{figure}

\subsection{Similarity Measurement and Clustering}
For NCTS data, we employ an attention-based neural network to measure the similarity between two segments. The network architecture and training process are the same as the attentive vector-to-sequence (Att-v2s) scoring in \cite{Lin2020}. The architecture of this transformer-based model consists of a multi-head self-attention module and several linear layers, as Figure \ref{fig:Att-v2s} shows. 

The input $\bm{m}_i$ is a sequence where each frame is a concatenation of two embeddings. Then the similarity matrix $\bm{S}$ is extracted as follows:
\begin{equation}
    \bm{S}_{i} = [S_{i1}, S_{i2}, ..., S_{in}] = f_{\mathrm{att}}(\bm{m}_{i})\\
\end{equation}
\begin{equation}
    \bm{m}_{i} = \left[\begin{matrix} \bm{x}_i & \bm{x}_i & ... & \bm{x}_i\\\bm{x}_1 &\bm{x}_2 & ... & \bm{x}_n\end{matrix}\right], 
\end{equation}
where $\bm{S}_{i}$ is the i-th row of the similarity matrix $\bm{S}$, $\bm{m}_{i}$ is the i-th row of the network input, and $\bm{x}_i$ is the speaker embedding of the i-th segment. For j-th entry $\bm{m}_{ij}=\left[\begin{matrix}\bm{x}_i\\\bm{x}_j\end{matrix}\right]$ in $\bm{m}_{i}$, the corresponding output is $S_{ij}$.

The first linear layer contains 256 units. The self-attention-based encoder contains two heads with 128 attention units. The dimension of the last two linear layers is 1024 and 1, followed by a sigmoid function. During the training phase, we employ the Diaconis augmentation \cite{li2019discriminative} on the embedding sequences with a probability of 0.5 for data augmentation. The binary cross-entropy (BCE) loss function and the stochastic gradient descent (SGD) optimizer are employed with an initial learning rate of 0.01. The learning rate decreases twice to 0.0001 with a factor of 0.1. Then, we finetune the model on the whole development set for 30 epochs with a fixed learning rate of 0.0001, and we do not use a validation set. For inference, we use the segments with a window length of 1.5s and a shift of 0.25s. Finally, we employ spectral clustering (SC) \cite{von2007tutorial} to get the diarization result. For more details, please refer to \cite{Lin2020}.

For CTS data, we use cosine distance to measure the similarity between two segments. Then, we perform AHC to cluster these segments. Note that this clustering step is not to obtain the final diarization result. We want to find the speech region for each speaker, and the speech region should contain as less overlap as possible. Since we already know that CTS data only contains two speakers, we can cluster all segments into two clusters, and the center of each cluster is the mean of all speaker embeddings. Since our purpose is to find two speakers' speeches without overlap, we set a high stop threshold of 0.6. And the two clusters will be used to extract the target speaker embedding for TSVAD later. 

After we get these two clusters, we can still assign other segments to these clusters to get the final diarization results for comparison. The center embedding for each cluster is fixed. Once the cosine distance between the speaker embedding of a segment and each cluster center embedding is lower than another predefined threshold, we consider it as an overlapping segment, and it will be added to each cluster. The threshold is set to 0.0, which is tuned from the development set.

In the next section, we will use these speech regions to extract the speaker embedding for each speaker and perform TSVAD to obtain a more accurate result. 

\begin{figure}
    \centering
    \includegraphics[scale=0.85]{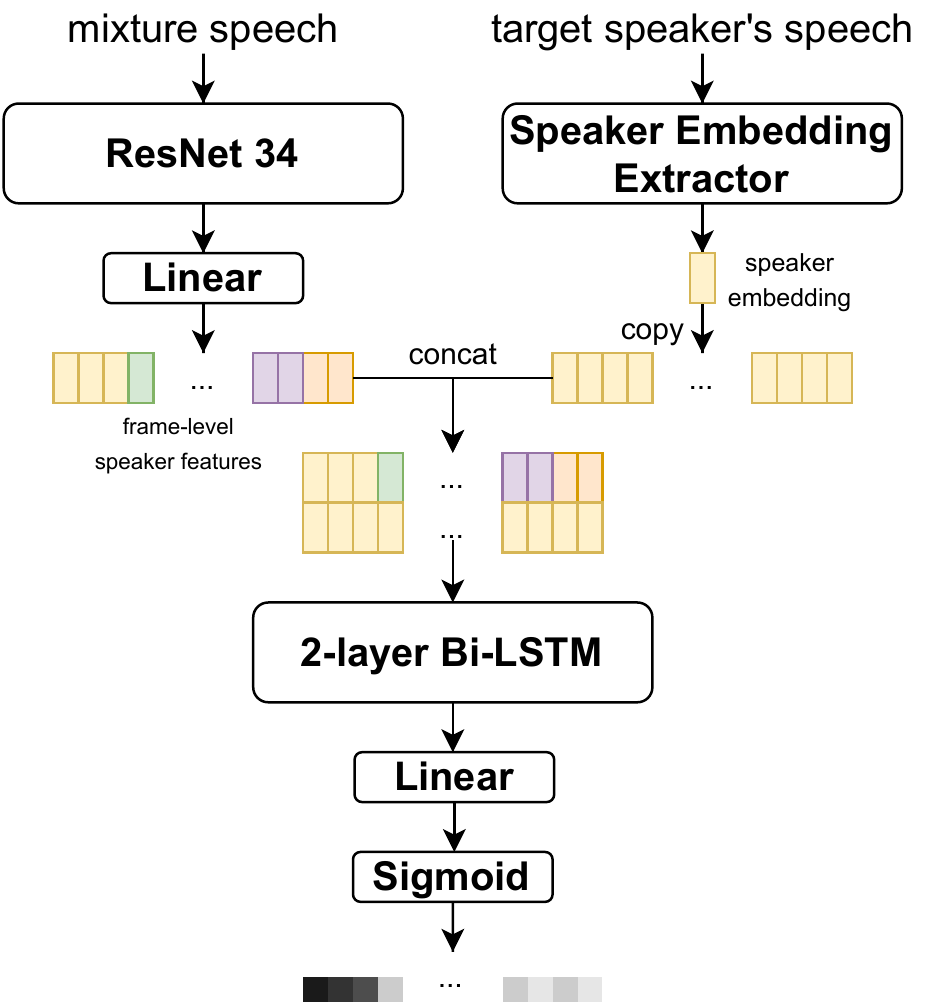}
    \caption{The architecture of the TSVAD model}
    \label{fig:TSVAD}
\end{figure}

\subsection{TSVAD}

We only perform TSVAD for CTS data. The model is similar to the model in \cite{ding2019personal, medennikov2020target}, but the training strategy and network architecture are different. First, we only detect the speech region, which means that the task becomes a binary classification. Another difference is that we use a ResNet34 to further extract the frame-level speaker identity information instead of directly using acoustic features. Figure \ref{fig:TSVAD} shows the structure of our TSVAD model. First, a ResNet with a fully-connected layer extracts the 128-dimensional frame-level speaker identity vector. Then, the target speaker embedding is concatenated to each frame of speaker identity vector as the input of 
a target speaker detector, which consists of two Bi-LSTM layers. Finally, a fully-connected layer predicts if a frame contains the target speaker. The speaker embedding extractor is the same as the model in Section 3.3. 

In the training phase, the parameters of ResNet34 is the same as the front-end ResNet of the speaker embedding extractor. These parameters of ResNet34 are frozen during the early training phase, and we only update the parameters in the Bi-LSTM and Linear layers. After these parameters converging, we unfreeze the parameters of the ResNet34 in the left of Figure 3. Then we train the whole model for several epochs until converging.

The acoustic features are 80-dimensional log Mel-filterbank energies with a frame length of 25ms and a hop size of 10ms. During the training phase, we first use Kaldi Tools \cite{Povey_ASRU2011} to get the VAD label for training data. We randomly select 8s of a speaker's audio to extract the target speaker embedding and randomly select 4s to 20s speech as the input of ResNet34 to extract frame-level speaker identity, as show in Figure \ref{fig:TSVAD}. The learning rate is set to 0.0001 when ResNet34 is frozen and 0.00001 when ResNet34 is unfrozen. The stochastic gradient descent (SGD) optimizer and the binary cross-entropy (BCE) loss are employed. During the finetuning stage, we use the first 41 recordings as the finetuning set and the remaining 20 recordings as the validation set. The learning rate is set to 0.00001. During the inference phase, some post-process are employed to get the final diarization result. First, we perform 11-tap median filtering on the output of the neural network. Then we decide the target speech for each speaker by a predefined threshold of 0.65. Note that we only perform TSVAD on the speech region. Thus, some frames may be misclassified as non-speech because the output of each target speaker is lower than the threshold. We choose the speaker with larger output as the target speaker for these frames.

\section{Experimental Results}
\subsection{VAD}
In the DIHARD III challenge, the VAD labels for the evaluation set are available in task 1 but should be excluded from task 2, so we do not test our model on the evaluation set. The model is trained on the 90\% of the development set and validated on the remaining 10\% data. Table \ref{tab:vad} shows that the training accuracy is 96.8\% and validation accuracy is 94.9\%, which means that our model is not overfitted. 

\begin{table}[htb]
  \caption{VAD accuracy on the development set}
  \label{tab:vad}
  \centering
  \begin{tabular}[c]{lcc}
    \toprule
    & \textbf{Training set} & \textbf{Validation set} \\
    \midrule
    Accuracy & 96.8\% &  94.9\% \\
    \bottomrule
  \end{tabular}
\end{table}

\subsection{Clustering}

Table \ref{tab:dev} shows the results of NCTS data, CTS data on the development set in task 1. For NCTS data, we provide the DER before finetuning on the development set. For CTS data, we provide the DER of AHC clustering on the development set. Note that all result of CTS data is on the last 20 recordings. For the whole dataset, the result is the combination of the CTS data and NCTS data on the development set. Table \ref{tab:eval} shows the total DER on the evaluation set.

\begin{table}[htb]
  \caption{System performance (DER) on development dataset (task 1)}
  \label{tab:dev}
  \centering
  \begin{tabular}[c]{lcc}
    \toprule
    \textbf{Dataset} & \textbf{Method} & \textbf{DER (\%)} \\
    \midrule
    NCTS & att-v2s + SC & 16.05 \\
    CTS & Cosine + AHC & 15.07 \\
    CTS & TSVAD & 10.60 \\ 
    CTS (adapt) & TSVAD round 1 & 7.80\\
    CTS (adapt) & TSVAD round 2 & 7.63 \\
    \bottomrule
  \end{tabular}
\end{table}

\subsection{TSVAD}
Table \ref{tab:dev} also shows the results of the TSVAD model on the CTS data. We first directly evaluate the performance of the CTS dataset. Then, we use the first 41 recordings in the CTS for finetuning (adapt) and test on the remaining 20 recordings. After obtaining the diarization results (round 1) from the TSVAD model, we extract each speaker's speech from this result and feed it to our TSVAD model again to get the results (round 2). Results show that the DER also decreases in the 2nd round, but it no longer changes in the later round. 

The results show that the TSVAD model significantly reduces the DER from 15.07\% to 7.63\%. Table \ref{tab:eval} shows the total DER with TSVAD on the evaluation set. Since we only submit our TSVAD-based results to task 2, some entries in Table \ref{tab:eval} are missed. 

\begin{table*}[!htbp]
  \caption{System performance (DER) on evaluation dataset (task 1 \& 2)}
  \label{tab:eval}
  \centering
  \begin{tabular}[c]{lcccc}
    \toprule
    & \textbf{Dataset} & \textbf{Method} & \textbf{DER on full set (\%)}  & \textbf{DER on core set (\%)} \\
    \midrule
    \multirow{2}{*}{task1} & NCTS (adapt) \& CTS & att-v2s + SC \& Cosine + AHC & 16.34 & 17.03 \\
    & NCTS (adapt) \& CTS (adapt) & att-v2s + SC \& TSVAD round 2& 13.39 & 15.43 \\ \midrule
    \multirow{2}{*}{task2} & NCTS (adapt) \& CTS & att-v2s + SC \& Cosine + AHC & - & - \\
    & NCTS (adapt) \& CTS (adapt) & att-v2s + SC \& TSVAD round 2& 18.90 & 21.63 \\
    \bottomrule
  \end{tabular}
\end{table*}

\subsection{Discussion}
Our TSVAD model shows a good performance on the CTS data. From the experiments in \cite{medennikov2020target}, it seems that xvector is not as good as the ivector as the target speaker's embedding. However, in our experiment, the speaker embedding extracted by the ResNet-based model shows good performance compared with the ivector-based method. The main reason may be that we use a ResNet34 to further extract the frame-level speaker feature, which can help the later Bi-LSTM layers to find the relationship between each frame and the target speaker's embedding. Besides, we copy the parameters of the pre-trained speaker embedding extract to the ResNet34 to extract the frame-level speaker identity, which speeds up the converging. 

Although TSVAD shows a good performance on the CTS data, it has poor performance on the NCTS data. The reason may be that the NCTS data comes from various domains and some recordings are extremely noisy. Our speaker embedding cannot extract enough speaker identity for such noisy recordings. We will train our model for each domain in the future. 

\section{Conclusion}
In this paper, we provide a detailed description of our diarization system. We break up the dataset into CTS and NCTS data and evaluate the performance of them. We also employ TSVAD for the CTS to find the overlapping region and reduce the DER. In task 1, our final submission achieves a DER of 13.39 and 15.43 on the full and core evaluation set. In task 2, we achieve a DER of 18.90 and 21.63 on the full and core evaluation set. We rank 4th place on task 1 and 3rd place on task 2. 
\bibliographystyle{IEEEtran}

\bibliography{refer}

\begin{thebibliography}{10}
\providecommand{\url}[1]{#1}
\csname url@samestyle\endcsname
\providecommand{\newblock}{\relax}
\providecommand{\bibinfo}[2]{#2}
\providecommand{\BIBentrySTDinterwordspacing}{\spaceskip=0pt\relax}
\providecommand{\BIBentryALTinterwordstretchfactor}{4}
\providecommand{\BIBentryALTinterwordspacing}{\spaceskip=\fontdimen2\font plus
\BIBentryALTinterwordstretchfactor\fontdimen3\font minus
  \fontdimen4\font\relax}
\providecommand{\BIBforeignlanguage}[2]{{%
\expandafter\ifx\csname l@#1\endcsname\relax
\typeout{** WARNING: IEEEtran.bst: No hyphenation pattern has been}%
\typeout{** loaded for the language `#1'. Using the pattern for}%
\typeout{** the default language instead.}%
\else
\language=\csname l@#1\endcsname
\fi
#2}}
\providecommand{\BIBdecl}{\relax}
\BIBdecl

\bibitem{ryanta2018enhancement}
N.~Ryanta, E.~Bergelson, K.~Church, A.~Cristia, J.~Du, S.~Ganapathy,
  S.~Khudanpur, D.~Kowalski, M.~Krishnamoorthy, R.~Kulshreshta \emph{et~al.},
  ``Enhancement and analysis of conversational speech: Jsalt 2017,'' in
  \emph{2018 IEEE International Conference on Acoustics, Speech and Signal
  Processing (ICASSP)}.\hskip 1em plus 0.5em minus 0.4em\relax IEEE, 2018, pp.
  5154--5158.

\bibitem{ryant2020third}
N.~Ryant, K.~Church, C.~Cieri, J.~Du, S.~Ganapathy, and M.~Liberman, ``Third
  dihard challenge evaluation plan,'' \emph{arXiv preprint arXiv:2006.05815},
  2020.

\bibitem{nagrani2020voxsrc}
A.~Nagrani, J.~S. Chung, J.~Huh, A.~Brown, E.~Coto, W.~Xie, M.~McLaren, D.~A.
  Reynolds, and A.~Zisserman, ``Voxsrc 2020: The second voxceleb speaker
  recognition challenge,'' \emph{arXiv preprint arXiv:2012.06867}, 2020.

\bibitem{xiao2020microsoft}
X.~Xiao, N.~Kanda, Z.~Chen, T.~Zhou, T.~Yoshioka, Y.~Zhao, G.~Liu, J.~Wu,
  J.~Li, and Y.~Gong, ``Microsoft speaker diarization system for the voxceleb
  speaker recognition challenge 2020,'' \emph{arXiv preprint arXiv:2010.11458},
  2020.

\bibitem{horiguchi2020end}
S.~Horiguchi, Y.~Fujita, S.~Watanabe, Y.~Xue, and K.~Nagamatsu, ``End-to-end
  speaker diarization for an unknown number of speakers with encoder-decoder
  based attractors,'' \emph{arXiv preprint arXiv:2005.09921}, 2020.

\bibitem{Lin2020}
\BIBentryALTinterwordspacing
Q.~Lin, Y.~Hou, and M.~Li, ``{Self-Attentive Similarity Measurement Strategies
  in Speaker Diarization},'' in \emph{Proc. Interspeech 2020}, 2020, pp.
  284--288. [Online]. Available:
  \url{http://dx.doi.org/10.21437/Interspeech.2020-1908}
\BIBentrySTDinterwordspacing

\bibitem{voxceleb}
A.~Nagrani, J.~Chung, and A.~Zisserman, ``Voxceleb: a large-scale speaker
  identification dataset,'' 2017.

\bibitem{mccowan2005ami}
I.~McCowan, J.~Carletta, W.~Kraaij, S.~Ashby, S.~Bourban, M.~Flynn,
  M.~Guillemot, T.~Hain, J.~Kadlec, V.~Karaiskos \emph{et~al.}, ``The ami
  meeting corpus,'' in \emph{Proceedings of the 5th International Conference on
  Methods and Techniques in Behavioral Research}, vol.~88.\hskip 1em plus 0.5em
  minus 0.4em\relax Citeseer, 2005, p. 100.

\bibitem{janin2003icsi}
A.~Janin, D.~Baron, J.~Edwards, D.~Ellis, D.~Gelbart, N.~Morgan, B.~Peskin,
  T.~Pfau, E.~Shriberg, A.~Stolcke \emph{et~al.}, ``The icsi meeting corpus,''
  in \emph{2003 IEEE International Conference on Acoustics, Speech, and Signal
  Processing, 2003. Proceedings.(ICASSP'03).}, vol.~1.\hskip 1em plus 0.5em
  minus 0.4em\relax IEEE, 2003, pp. I--I.

\bibitem{musan}
D.~Snyder, G.~Chen, and D.~Povey, ``{MUSAN}: {A} {Music}, {Speech}, and {Noise}
  {Corpus},'' \emph{arXiv:1510.08484}, 2015.

\bibitem{he2016deep}
K.~He, X.~Zhang, S.~Ren, and J.~Sun, ``Deep residual learning for image
  recognition,'' in \emph{Proceedings of the IEEE conference on computer vision
  and pattern recognition}, 2016, pp. 770--778.

\bibitem{cai2018exploring}
W.~Cai, J.~Chen, and M.~Li, ``Exploring the encoding layer and loss function in
  end-to-end speaker and language recognition system,'' \emph{arXiv preprint
  arXiv:1804.05160}, 2018.

\bibitem{deng2019arcface}
J.~Deng, J.~Guo, N.~Xue, and S.~Zafeiriou, ``Arcface: Additive angular margin
  loss for deep face recognition,'' in \emph{Proceedings of the IEEE/CVF
  Conference on Computer Vision and Pattern Recognition}, 2019, pp. 4690--4699.

\bibitem{qin2020ffsvc}
X.~Qin, M.~Li, H.~Bu, R.~K. Das, W.~Rao, S.~Narayanan, and H.~Li, ``The ffsvc
  2020 evaluation plan,'' \emph{arXiv preprint arXiv:2002.00387}, 2020.

\bibitem{li2019discriminative}
Q.~Li, F.~L. Kreyssig, C.~Zhang, and P.~C. Woodland, ``Discriminative neural
  clustering for speaker diarisation,'' \emph{arXiv preprint arXiv:1910.09703},
  2019.

\bibitem{von2007tutorial}
U.~Von~Luxburg, ``A tutorial on spectral clustering,'' \emph{Statistics and
  computing}, vol.~17, no.~4, pp. 395--416, 2007.

\bibitem{ding2019personal}
S.~Ding, Q.~Wang, S.-y. Chang, L.~Wan, and I.~L. Moreno, ``Personal vad:
  Speaker-conditioned voice activity detection,'' \emph{arXiv preprint
  arXiv:1908.04284}, 2019.

\bibitem{medennikov2020target}
I.~Medennikov, M.~Korenevsky, T.~Prisyach, Y.~Khokhlov, M.~Korenevskaya,
  I.~Sorokin, T.~Timofeeva, A.~Mitrofanov, A.~Andrusenko, I.~Podluzhny
  \emph{et~al.}, ``Target-speaker voice activity detection: a novel approach
  for multi-speaker diarization in a dinner party scenario,'' \emph{arXiv
  preprint arXiv:2005.07272}, 2020.

\bibitem{Povey_ASRU2011}
D.~Povey, A.~Ghoshal, G.~Boulianne, L.~Burget, O.~Glembek, N.~Goel,
  M.~Hannemann, P.~Motlicek, Y.~Qian, P.~Schwarz, J.~Silovsky, G.~Stemmer, and
  K.~Vesely, ``The kaldi speech recognition toolkit,'' in \emph{IEEE 2011
  Workshop on Automatic Speech Recognition and Understanding}.\hskip 1em plus
  0.5em minus 0.4em\relax IEEE Signal Processing Society, Dec. 2011, iEEE
  Catalog No.: CFP11SRW-USB.

\end{thebibliography}

\end{document}